\input phyzzx
\input maggieref
\newcount\mongocount
\mongocount=1
\def\Figure#1#2#3{
      \vbox to #3in{\hsize=#2in
        \vfil
         \includegraphics{#1}
    }
}
\def\figcap#1#2{
\vtop{\tenpoint\singlespace
\hsize=#1in\smallskip\noindent Figure\ \ \the\mongocount.\ \  #2
\global\advance\mongocount by 1\bigskip}}
\def\mongofigure#1#2#3#4#5{\centerline{\Figure{#1}{#2}{#3}
\figcap{#4}{#5}}}

\hoffset=0.375in
\overfullrule=0pt

\def\eff{{\rm eff}}
\def\psf{{\rm psf}}
\def\sat{{\rm sat}}

\def\psf{{\rm psf}}
\def\dol{{D_{\rm l}}}
\def\dls{{D_{\rm ls}}}
\def\dos{{D_{\rm s}}}

\def\max{{\rm max}}

\def\kpc{{\rm kpc}}
\def\pc{{\rm pc}}

\def\bx{{\bf x}}
\def\bk{{\bf k}}

\twelvepoint
\font\bigfont=cmr17
\centerline{\bigfont Microlensing and the Stellar Mass Function}
\bigskip
\centerline{{\bf Andrew Gould}\footnote{1}{Alfred P.\ Sloan Foundation Fellow}}
\smallskip
\centerline{Dept of Astronomy, Ohio State University, Columbus, OH 43210}
\smallskip
\centerline{e-mail gould@payne.mps.ohio-state.edu}
\bigskip
\singlespace
\centerline{\bf Abstract}

Traditional approaches to measuring the stellar mass function (MF) are
fundamentally limited because objects are detected based on their luminosity,
not their mass.  These methods are thereby restricted to luminous and 
relatively nearby stellar populations.  Gravitational microlensing promises to 
revolutionize our understanding of the MF.  It is already technologically 
feasible to measure the MFs of the Galactic disk and Galactic bulge as 
functions of position, although the actual execution of this program requires 
aggressive ground-based observations including infrared interferometry, as 
well as the launching of a small satellite telescope.  Rapid developments in 
microlensing, including the new technique of ``pixel lensing'' of unresolved 
stars, will allow one to probe the MF and luminosity function of nearby 
galaxies.  Such observations of M31 are already underway, and pixel-lensing 
observations of M87 with the {\it Hubble Space Telescope} would permit 
detection of dark intra-cluster objects in Virgo.  Microlensing techniques can 
also be applied to investigate the star-formation history of the universe and 
to search for planets with masses as small as the Earth's.

Based on an invited talk at the January 1996 AAS meeting in San Antonio.

PASP (June 1996) in press, (c) ASP, reproduced with permission.

\singlespace 

\bigskip
\chapter{Introduction}

	Few would argue with the assertion that the best way to find out 
about stars is to observe them.  However, while this traditional approach
has been and continues to be very successful, it does have its limitations.
To be observed, a star must be luminous enough and close enough to be
detected.  This obviously creates a heavy selection bias toward brighter
stars.  It is generally possible to quantify and compensate for this bias.
Nevertheless, as stars get sufficiently faint, the available volume becomes
so small that the statistics are compromised.  Below the hydrogen-burning
limit ($M\sim 0.08\,M_\odot$), the luminosity of the star (or rather brown 
dwarf) depends 
not only on its mass but also its age.  The age in turn is strongly
correlated with position; younger objects are confined close to the Galactic
plane and older ones are distributed at greater heights.  For these reasons,
our ability to probe the stellar mass function (MF) breaks down rapidly
as one approaches and then crosses the hydrogen-burning limit.

	Other problems emerge if we try to measure the MF at locations
other than the immediate solar neighborhood.  First, of course, it becomes
increasingly difficult to detect faint stars.  However, there is a more
fundamental problem.  The masses of individual stars are estimated with the
help of a mass-luminosity relation, where the luminosity is given in some
observed band such as $V$ or $K$. 
This relation is in turn calibrated using stars in (e.g.\ visual)
binaries whose masses are well determined.  Since the distances (and so 
absolute
magnitudes) are also well determined, the calibration is purely empirical.
Unfortunately, it is also purely local.  The mass-luminosity relation may
depend on location through metallicity, age, or other factors.  
It is impossible to measure the MF in a region far
from the solar neighborhood without a direct measurement of the mass of at 
least some stars in that region.

	Microlensing offers another approach.  Microlensing occurs when
an object (``the lens'') comes close to the line of sight between the
observer and a distant star (``the source'').  The gravity of the lens
deflects and thereby magnifies the light from the source.  As the
lens moves closer and then farther from the line of sight, the magnification
rises from unity to some maximum value and then declines symmetrically.
This characteristic one-time variation is then the signature of microlensing.

	The major advantage of microlensing over traditional methods of
measuring the MF is obvious:  the microlensing effect is produced 
by the mass of the lens, not its luminosity.  Hence, microlensing-selected
samples are completely free of bias induced by the hydrogen-burning limit.  
Moreover, microlenses need not be nearby.  The method can be used to
probe the MF in remote regions of the Milky Way and even in other galaxies.

	On the other hand, it not at all obvious how one might go about
measuring the mass of a lens once it is detected.  Microlensing experiments
were initially suggested as a method to detect Massive Compact Halo Objects
(MACHOs) in the Milky Way halo by observing millions of source stars in the
Large Magellanic Cloud (LMC) (Paczy\'nski 1986).  They were not expected
to yield information about the mass of individual lenses to better 
than a factor $\sim 3$.  Rather, the goal was to measure the total mass in 
MACHOs and to make a rough estimate of the MACHO mass scale.  Two groups
have undertaken MACHO searches, MACHO (Alcock et al.\ 1993, 1995b, 1995d, 
1996a) and EROS (Aubourg et al.\ 1993, 1995; Ansari 1996).  

	However, several developments have now combined to
open the way to microlensing measurements of the MF.  
Paczy\'nski (1991) and Griest et al.\ (1991) independently suggested
that microlensing experiments toward the Galactic bulge could be used
to probe stars along the line of sight in the Galactic disk.  The 
OGLE (Udalski et al.\ 1994a, 1994b), MACHO (Alcock et al.\ 1995a, 1996b), 
and DUO
(Alard, Mao, \& Guibert 1995; Alard 1996) collaborations have all initiated 
such searches and the EROS collaboration plans to join them soon.  A new
Japan-New Zealand collaboration MOA is also actively preparing for
observations.
Kiraga \& Paczy\'nski (1994) pointed out that these searches are sensitive to
stars in the Galactic bulge as well as the disk.  Theoretical work by many
people has shown that it is possible to recover far more information about
individual masses than was originally believed.  Crotts (1992) and 
Baillon et al.\ (1993) developed ``pixel lensing'' techniques for finding 
lensing events of unresolved stars and applied them to observations of M31, 
thus creating the possibility of measuring the MF in external galaxies.
Several of the microlensing collaborations have developed the ability to
recognize ongoing events in real time.  These ``alert'' events are now
being followed by at least two world-wide networks, PLANET (Albrow et al.\ 
1996)
and GMAN (Pratt et al.\ 1996) with the aim of finding several classes
of higher order effects predicted by theorists, including signatures of
planets.

	In brief, microlensing research is proceeding at an explosive pace.
Paczy\'nski (1996) and Roulet \& Mollerach (1996) have given comprehensive
reviews of these developments.  Schneider, Ehlers, \& Falco (1992) place
microlensing in the broad context of other forms of gravitational lensing.
Here the focus is on what microlensing can
tell us about stars.

\chapter{{\it HST} Star Counts}

\topinsert
\mongofigure{ps.lfreidqp}{6.4}{5.5}{6.4}{
Luminosity function of disk stars as determined from {\it HST} data by
Gould et al.\ (1996) (triangles) compared to several other determinations.
Shown are the ground-based photometric survey of Stobie et al.\ (1989)
(open circles), the parallax-star survey of Wielen et al.\ (1983)
(filled circles) and the parallax-star study of Reid et al.\ (1995) excluding
(filled squares) and including (open squares) the secondaries in binaries.
Error bars are shown only for the {\it HST} data to avoid cluttering.  Note
that the photometric studies of Gould et al.\ and Stobie et al.\ are in almost
perfect agreement.  Also note from the Reid et al.\ data that about half
of faint M stars are secondaries. 
From Gould et al.\ (1996).  Copyright American Astronomical Society,
reproduced with permission.
}\endinsert
\FIG\one{
Luminosity function of disk stars as determined from {\it HST} data by
Gould et al.\ (1996) (triangles) compared to several other determinations.
Shown are the ground-based photometric survey of Stobie et al.\ (1989)
(open circles), the parallax-star survey of Wielen et al.\ (1983)
(filled circles) and the parallax-star study of Reid et al.\ (1995) excluding
(filled squares) and including (open squares) the secondaries in binaries.
Error bars are shown only for the {\it HST} data to avoid cluttering.  Note
that the photometric studies of Gould et al.\ and Stobie et al.\ are in almost
perfect agreement.  Also note from the Reid et al.\ data that about half
of faint M stars are secondaries. 
From Gould et al.\ (1996).  Copyright American Astronomical Society,
reproduced with permission.
}

	Star counts with the {\it Hubble Space Telescope (HST)} 
(Bahcall et al.\ 1994; Gould, Bahcall \& Flynn 1996) provide a good 
illustration of both the recent progress and fundamental limitations of
traditional techniques.  On the one hand, {\it HST} observations allow one to
measure the luminosity function (LF) of M dwarfs up to several kpc from the
Sun.  These data have helped resolve a major conflict about the faint
end of the LF which had persisted for more than 10 years.  Photometric
surveys of stars within $\sim 100\,$pc (e.g.\ Stobie, Ishida, \& Peacock 1989)
have tended to show a sharply falling LF for stars $M_V>12$, while in their
more nearby parallax-based survey, Wielen, Jahreiss, \& Kr\"uger (1983) 
find a nearly flat LF.  

	There are important differences between the two types of surveys.
In photometric surveys, one estimates the absolute magnitude of the star
from the measured color and a local, empirically calibrated color-magnitude
relation.  The distance is then determined by comparing the absolute
and apparent magnitudes.  This method creates a statistical bias toward
including brighter-than-average stars in the sample, called Malmquist bias,
because the brighter stars can be detected over a larger volume.  It is
possible to correct for Malmquist bias, but this correction can be uncertain.
In parallax surveys, by contrast, the distance is known from the parallax,
and so the absolute magnitude is determined rather than estimated.  On the
other hand, of course, parallax surveys are limited statistically by the
small volume probed.

	The {\it HST} photometric survey has a mean
limiting magnitude of $I=23.7$, about 100 times fainter than is typical of
ground-based surveys.  It agrees almost perfectly with
the Stobie et al.\ photometric survey.  This shows first that the LF at
a few kpc from the Sun is the same as the LF at $\sim 100$ pc.  Moreover,
since the {\it HST} observations reach to the ``top'' of the Galactic
disk, they require only a modest correction for Malmquist bias.  Hence their
agreement with the more Malmquist-bias prone local sample is reassuring.
The conflict between the photometric and parallax LFs has been significantly
softened by the work of Reid, Hawley, \& Gizis (1995), who made substantial
new observations of parallax stars and did a thorough reanalysis of the data.
They find that if one includes only single stars and the primaries of
binaries (which is essentially all that photometric surveys detect), there
is very good agreement between the photometric and parallax surveys.
If the secondaries of binaries are re-included, then the luminosity function
still falls, but not as rapidly as the photometric surveys would indicate
(see Fig.\ \one).  	Thus, substantial progress has been made on the LF on 
several fronts.  

\topinsert
\mongofigure{ps.massqp}{6.4}{5.5}{6.4}{
Mass function derived from luminosity functions of Gould et al.\ (1996) for
$M_V>8$ (triangles) and of Wielen et al.\ (1983) for $M_V<8$ as shown in
Fig.\ \one.  Curve is a parabolic fit to the data excluding the last point.
From Gould et al.\ (1996).  Copyright American Astronomical Society,
reproduced with permission.
}\endinsert
\FIG\two{
Mass function derived from luminosity functions of Gould et al.\ (1996) for
$M_V>8$ (triangles) and of Wielen et al.\ (1983) for $M_V<8$ as shown in
Fig.\ \one.  Curve is a parabolic fit to the data excluding the last point.
From Gould et al.\ (1996).  Copyright American Astronomical Society,
reproduced with permission.
}

The MF constructed by Gould et al.\ (1996) from the {\it HST} LF
and the mass-luminosity relation of Henry \& McCarthy (1993) falls for
$M<0.45\, M_\odot$ ($M< 0.23\,M_\odot$ if number density is measured per unit
mass instead of log-mass).  
See Figure \two.  The {\it HST} counts yield a
number of other important results as well.  The vertical distribution of 
M stars
is measured for the first time.  The intermediate component (``thick disk'')
has a scale height $h\sim 660\,$pc, much smaller than usually assumed.
The thin disk is inconsistent with an exponential profile (which is usually
assumed) and is much closer to a sech${}^2$ law.  The total column density
of M stars is only $\sim 13\,M_\odot\,\pc^{-2}$, implying that the observed
column density of the disk is only $\Sigma\sim 40\,M_\odot\,\pc^{-2}$, about
20\% less than the best previous estimate.  Thus, the {\it HST} star counts
provide a good example (certainly not the only one) of continuing advances
in the study of stars by traditional techniques.  Bessel \& Stringfellow (1993)
give a good recent review of the traditional approaches to this problem.
See also Scalo (1986).

	On the other hand, Figures \one\ and \two\ 
also make clear the limitations
of these techniques.  First, of course, the graphs simply end at the hydrogen
burning limit.  Even the approach to the hydrogen limit is unclear.  The
{\it HST} data appear to show a rise at the last point, but this is based
on only six stars and so is not significant.  There are several intriguing
lines of argument that the MF is rising at this point, including circumstantial
evidence that young (and so more luminous) brown dwarfs are being detected and
mistaken for faint stars (Reid, Tinney, \& Mould 1994; Kirkpatrick et al.\
1994).  However, these results are based on only 17 detections and are
open to conflicting interpretations.  In brief, new approaches are needed
to supplement the old.

\chapter{Simple Microlensing}
\bigskip

	Consider a point mass $M$ and a point source of light which are
approximately aligned at distances $\dol$ and $\dos$ from the observer.
General relativity predicts that, in the weak field limit, the mass
(the ``lens'') deflects light by an angle $\alpha$,
$$\alpha = {4 G M\over b c^2},\eqn\alphadef$$
where $b$ is the impact parameter (distance of closest approach) and where
it is assumed that $b\ll\dol$, $b\ll\dls$.
Here $\dls\equiv\dos-\dol$ is the distance from the lens to the source.    
The source will then be split
into two images whose positions are governed by the equation
$$\theta_S^2 - \theta_I\theta_S + \theta_e^2 = 0,\eqn\lenseq$$
where $\theta_S$ and $\theta_I$ are the angular separations between the lens
and the source and image respectively, and where $\theta_e$ is the
angular Einstein radius,
$$\theta_e\equiv \sqrt{4 G M \dls\over c^2\dol\dos}.\eqn\thetae$$
  Equation \lenseq\ has two solutions:
$$\theta_{I\pm} = \pm x_\pm \theta_e\eqn\lenssol$$
where
$$x_\pm\equiv {\sqrt{x^2 + 4}\pm x\over 2},\qquad x\equiv 
{\theta_S\over\theta_e}.\eqn\xpdef$$
The $(\pm)$ symbol indicates that two images are on opposite sides of the lens.
The $(+)$ image has the larger separation and is on the same side as the
source.  
\topinsert
\mongofigure{ps.lensdiaghan}{6.8}{6.5}{6.8}{
Geometry of Microlensing.  Shown in (a) is the idealized case when
the observer (O), the lens (L), and source (S) are perfectly aligned.  
Light from S is
bent by an angle $\alpha=4 G M/r_e c^2$ [cf.\ eq.\ \alphadef] by the mass
$M$ of the lens.  The image (I) is deflected by angle $\theta_e$ relative to 
the true source position.  On the right the geometry is shown from the point
of view of the observer.  Again, the lens (solid circle) and source 
(small open circle)
are perfectly aligned.  The image is a ring of light (the ``Einstein Ring'')
between the two large non-bold circles which  
straddle the bold circle of angular radius $\theta_e$.  
Shown in (b) is the more generic case where
S and L are separated by a finite angle $\theta_S$.  There are now two
different deflection angles $\alpha_1$ and $\alpha_2$, one for each of the two
images $I_1$ and $I_2$.  The images no longer make a ring, but it is
still convenient to draw an imaginary Einstein ring (right).  The brighter 
image falls outside the ring and the fainter image falls inside.  Note that
$x$ is the separation of S and L in units of the Einstein ring.  Shown in
(c) is the magnification.  The images have the same surface brightness as
the source.  Assuming the latter is uniform, the magnification is then simply
the ratio of areas.  Unless the lens transits the source, the magnification
is given analytically by eq.\ (3.7).
}\endinsert
\FIG\three{
Geometry of Microlensing.  Shown in (a) is the idealized case when
the observer (O), the lens (L), and source (S) are perfectly aligned.  
Light from S is
bent by an angle $\alpha=4 G M/r_e c^2$ [cf.\ eq.\ \alphadef] by the mass
$M$ of the lens.  The image (I) is deflected by angle $\theta_e$ relative to 
the true source position.  On the right the geometry is shown from the point
of view of the observer.  Again, the lens (solid circle) and source 
(small open circle)
are perfectly aligned.  The image is a ring of light (the ``Einstein Ring'')
between the two large non-bold circles which  
straddle the bold circle of angular radius $\theta_e$.  
Shown in (b) is the more generic case where
S and L are separated by a finite angle $\theta_S$.  There are now two
different deflection angles $\alpha_1$ and $\alpha_2$, one for each of the two
images $I_1$ and $I_2$.  The images no longer make a ring, but it is
still convenient to draw an imaginary Einstein ring (right).  The brighter 
image falls outside the ring and the fainter image falls inside.  Note that
$x$ is the separation of S and L in units of the Einstein ring.  Shown in
(c) is the magnification.  The images have the same surface brightness as
the source.  Assuming the latter is uniform, the magnification is then simply
the ratio of areas.  Unless the lens transits the source, the magnification
is given analytically by eq.\ (3.7).
}

Figure \three(a) illustrates this geometry for the special case when the
lens is perfectly aligned with the source.  
Figure \three(b) shows the more
typical case when the alignment is close, but not perfect.

	Gravitational lenses, like telescope lenses, produce images which
transform the size and shape of the source, but not the surface intensity
of its light (Liouville 1837; Misner, Thorne, \& Wheeler 1973).  Thus
for a source with uniform surface brightness, the magnification is given by 
the ratio of the areas of the images to the area of the source.  
This is shown in Figure \three(c).  Each point of light from the source
is stretched when it is mapped onto the images.  From Figure \three(b) one
sees that it is stretched by a factor $x_\pm/x$ in the tangential direction,
and by $d x_\pm /d x$ in the radial direction.
Hence the two images of a point source are magnified by,
$$A_\pm(x) = \Bigg|{x_\pm\over x}\,{d x_\pm\over d x} \Bigg| =
{x_\pm^2\over x_+^2 - x_-^2},\eqn\apm$$
yielding a total magnification,
$$A(x) = A_+ + A_- ={x^2 + 2\over x\sqrt{x^2+4}}.\eqn\aofx$$
Unless the lens transits (or nearly transits) the source (see \S\ 5.2), 
the magnification of a finite source is also very well approximated by 
equation \aofx.
For typical microlensing events, $\theta_e\lsim 1\,$mas, which is not
generally detectable.  The magnification is therefore the only signature
that a lensing event is in progress.  However, as explained below,
the magnification is not 
directly observable.  What can be observed is the changing magnification
as the lens moves on a straight line with projected separation from the
source in units of the Einstein
ring 
$$x(t;t_0,\beta,\omega) = \sqrt{\omega^2(t-t_0)^2 + \beta^2}.\eqn\xoft$$
Here $t_0$ is the time of closest approach, $\beta$ is the impact parameter
in units of $\theta_e$, 
$$\omega^{-1} \equiv t_e = {r_e\over v} = {\dol\theta_e\over v}
,\eqn\omegadef$$
is the characteristic time scale of the event, $r_e\equiv \dol\theta_e$ is the
physical Einstein radius, and $v$ is the speed of
the lens relative to the Earth-source line of sight.

	What is actually observed is the change in the flux of the source
star
$$F(t;t_0,\beta,\omega,F_0) = \{A[x(t;t_0,\beta,\omega)]-1\}F_0,\eqn\foft$$
where $F_0$ is the unlensed flux of the  star.  (Note that $F_0$ is not known
{\it a priori} from the pre-event light curve.  What is measured before and
after the event is
$F_0+B$ where $B$ is the sum of the fluxes from other non-lensed objects
such as a binary companion, the lens itself, or random field stars.  For most
lensing events observed to date, the quality of the data has not been high
enough to measure $F_0$ and $B$ separately.  Rather, the measured value
of $F_0+B$ is combined with a statistical estimate of $B$ to obtain an
estimate of $F_0$.  This statistical approach is certainly adequate
for estimating the total density of dark objects, and blending is only a
minor uncertainty in the results of MACHO and EROS.  However, blending is
a much more important problem when estimating the MF.  See \S\ 4.
Direct measurements of $F_0$ which are free from blending corrections
will be made in the future
when high-quality data will be obtained from follow-up photometry.
See \S\ 5.2.)

	Equation \foft\ immediately reveals the fundamental shortcoming
of microlensing: even if the light curve is observed with perfect accuracy,
it yields only four parameters: $t_0$, $\beta$, $\omega$, and $F_0$.  Of these
four, only $\omega$ contains any information about the lens.  Moreover,
it is clear from equation \omegadef\ that $\omega$ is a complicated
combination of the quantities one would like to know.

	Faced with this difficulty, there are two options for using 
microlensing to learn about the MF of stars.  First, make a statistical
analysis of the measured time scales.  As discussed in the next section, this
approach requires assumptions about the velocity and space distributions of
both the lenses and sources. Second, attempt to get more information
so as to be able to measure the masses of individual lenses.

\chapter{Initial Measurements of the Mass Function}

	Two groups have used the measured time scales of the first 51 
events detected by MACHO
and OGLE toward the bulge to attempt to determine the MF of the lensing
stars along the line of sight in the disk and in the bulge itself
(Zhao, Spergel, \& Rich 1995; Han \& Gould 1996a).  See also Jetzer (1994).
The results are subject
to significant uncertainties, but are nonetheless striking.  Han \& Gould
(1996a) considered three families of MFs: power law with cut off, Gaussian,
and Hubble.  The Hubble is the MF determined from the {\it HST} data
by Gould et al.\ (1996), augmented by the observed density of white
dwarfs.  
\topinsert
\mongofigure{ps.han}{6.4}{5.5}{6.4}{
Best fit Gaussian and power-law mass functions determined by Han \& Gould 
(1996a) from the first 51 microlensing events reported by MACHO and OGLE 
from observations toward the Galactic bulge.  Also shown is the Hubble
mass function as determined by Gould et al.\ (1996) from {\it HST} star
counts (see Fig.\ \two) augmented to include the observed density of white
dwarfs.  The power law is favored over the Hubble at the $5.5\,\sigma$
level and over the Gausssian at the $3\,\sigma$ level.  
From Han \& Gould (1996a).  Copyright American Astronomical Society,
reproduced with permission.
}\endinsert
\FIG\six{
Best fit Gaussian and power-law mass functions determined by Han \& Gould 
(1996a) from the first 51 microlensing events reported by MACHO and OGLE 
from observations toward the Galactic bulge.  Also shown is the Hubble
mass function as determined by Gould et al.\ (1996) from {\it HST} star
counts (see Fig.\ \two) augmented to include the observed density of white
dwarfs.  The power law is favored over the Hubble at the $5.5\,\sigma$
level and over the Gausssian at the $3\,\sigma$ level.  
From Han \& Gould (1996a).  Copyright American Astronomical Society,
reproduced with permission.
}
The best fits are shown in Figure \six.  The best fit power law,
with a near-Saltpeter slope of 2.1 and lower-mass cut off at $M=0.04\,M_\odot$,
is preferred over the Hubble MF at $>5\,\sigma$.  (More formally, the $\chi^2$
of the Hubble MF is greater than that of the best power law by
$\chi^2_{\rm PL}-\chi^2_{HST} = 5.5^2$.)\ \ The primary reason for the poor
fit of the Hubble MF is that it fails to reproduce the short events.
Thus, microlensing seems
to be telling us that there are more low mass objects in either the disk or
the bulge than are observed locally.  One might be concerned that the
Hubble MF does not include the binary companions catalogued by Reid et al.\
(1995), but it is clear from a comparison between Figures \one\ and \six\ 
that this correction would not make a significant difference.

	However, there are three problems with this approach which lessen
one's confidence in the results.  First, MACHO and OGLE determined the
time scales assuming that there is no blending 
[i.e., $B=0$ in the discussion below eq.\ \foft].  If the events were in fact
significantly blended, the true time scales would be longer than reported.
Since the mass estimate scales $M\propto t_e^2$, the lower-mass cut off
might be significantly underestimated.  This problem will likely disappear
as more events accumulate, particularly giant-source events (Gould 1995d)
which are much less susceptible to blending.  Second, the results are wholly
dependent on the assumptions  adopted for the velocity and spatial 
distributions of the lenses and sources.  While the source distributions
are reasonably well constrained by observations, the distributions of lenses
must be inferred from the distributions of more luminous material.  In fact,
Zhao et al.\ (1995) used somewhat different distributions from those of
Han \& Gould (1996a) and got somewhat different results.  However, the
principal problem is that the mass of any given lens is very poorly
constrained, so that (given the uncertainty in the lens velocities and
positions) one will never obtain more than a crude picture of the mass
spectrum, regardless of statistics.  See for example the Gaussian versus
power-law fits in Figure \six.  If the preliminary detection of a large
population of unseen low-mass objects is confirmed by continuing observations,
then even such a crude determination would be quite important.  However,
the real interest would still be in obtaining detailed MFs for
the disk and bulge separately.

\chapter{Mass Measurements of Individual Microlenses}
\topinsert
\mongofigure{ps.parpasp}{6.8}{6.5}{6.8}{
Geometry of microlensing parallax.  Part (a) is an elaboration
of the lensing geometry for a hypothetical observer
perfectly aligned with the source-lens axis.  As in Fig.\ \three(a),
$\alpha$ is the bending angle and $\theta_e$ is the angular Einstein radius.
The physical Einstein ring is the circle centered on L.  Its radius, $r_e$,
is not explicitly labeled.  In addition, the diagram shows $\tilde r_e$,
the projection of $r_e$ onto the plane of the observer.  Parts (b)
and (c) show how $\tilde r_e$ is measured by parallax.  Part (b) shows
the source position in the Einstein ring as a function
of time (labeled in days) as seen from the Earth and satellite.  
The vector separation $\Delta{\bf x}=
(\omega\Delta t,\Delta\beta)$ remains a constant during the event.  The
resulting light curves as seen from the Earth and satellite are shown in
(c).  By measuring the impact parameters $\beta$ and $\beta'$ and the
times of maximum $t_0$ and $t_0'$, one can reconstruct the geometry of (b)
and so determine $\Delta \bf x$.   From (a) it is clear that the magnitude
of this vector, $\Delta x$, 
is the distance to the satellite $d_{sat}$ as a fraction of $\tilde r_e$.
One therefore recovers $\tilde r_e = d_{sat}/\Delta x$.  The solution
is actually degenerate because from the light curves (c) alone one cannot
tell whether the source trajectories in (b) pass on the same or opposite
sides of the lens.  As discussed in the text, the degeneracy can be broken.
Parts (b) and (c) adapted from Gould (1994b).  
Copyright American Astronomical Society, reproduced with permission.
}\endinsert
\FIG\seven{
Geometry of microlensing parallax.  Part (a) is an elaboration
of the lensing geometry for a hypothetical observer
perfectly aligned with the source-lens axis.  As in Fig.\ \three(a),
$\alpha$ is the bending angle and $\theta_e$ is the angular Einstein radius.
The physical Einstein ring is the circle centered on L.  Its radius, $r_e$,
is not explicitly labeled.  In addition, the diagram shows $\tilde r_e$,
the projection of $r_e$ onto the plane of the observer.  Parts (b)
and (c) show how $\tilde r_e$ is measured by parallax.  Part (b) shows
the source position in the Einstein ring as a function
of time (labeled in days) as seen from the Earth and satellite.  
The vector separation $\Delta{\bf x}=
(\omega\Delta t,\Delta\beta)$ remains a constant during the event.  The
resulting light curves as seen from the Earth and satellite are shown in
(c).  By measuring the impact parameters $\beta$ and $\beta'$ and the
times of maximum $t_0$ and $t_0'$, one can reconstruct the geometry of (b)
and so determine $\Delta \bf x$.   From (a) it is clear that the magnitude
of this vector, $\Delta x$, 
is the distance to the satellite $d_{sat}$ as a fraction of $\tilde r_e$.
One therefore recovers $\tilde r_e = d_{sat}/\Delta x$.  The solution
is actually degenerate because from the light curves (c) alone one cannot
tell whether the source trajectories in (b) pass on the same or opposite
sides of the lens.  As discussed in the text, the degeneracy can be broken.
Parts (b) and (c) adapted from Gould (1994b).  
Copyright American Astronomical Society, reproduced with permission.
}
	Figure \seven(a) is an elaboration of Figure \three(a), 
the Einstein ring 
diagram for a hypothetical observer perfectly aligned with the
lens-source axis.
The additional feature of this figure is $\tilde r_e$, the projection
of the Einstein radius $r_e$ onto the observer plane.   The algebraic
expression for the projected radius is $\tilde r_e= (\dos/\dls) r_e$,
but its physical significance can be understood directly from Figure 
\seven(a)
which makes clear that knowledge of $\tilde r_e$, $\theta_e$, and $\dos$
would fix the distance to the lens, $\dol$.  Then, from equation
\thetae\ one could determine $M$, and from equation \omegadef\ and the
measured value of the time scale, $t_e$, one could determine $v$.

	In fact, closer inspection of Figure \seven(a) shows that $M$ can be 
determined from $\tilde r_e$ and $\theta_e$, even if $\dos$ is unknown.
Using the small angle approximation (certainly valid for angles $\lsim$mas),
one sees from the diagram that $\alpha:\tilde r_e = \theta_e:r_e$, implying 
$$ \theta_e\tilde r_e = \alpha r_e = {4 G\over c^2}\,M,\eqn\massdet$$
where in the last step I have made use of the Einstein light deflection
equation \alphadef.

	Is it possible to measure the $\tilde r_e$ and $\theta_e$ for
a representative sample of events and so determine the MF?  With some work,
yes.  Moreover, since $\dos$ is typically known to $\sim 10\%$ for bulge
sources, not only the mass, but also the approximate distance would be known.
Hence one could determine the bulge and disk MFs separately.

\section{Parallax Satellite Measurements of $\tilde r_e$}

	For galactic bulge sources at $\dos\sim R_0=8\,\kpc$,
$\tilde r_e= 8\sqrt{(M/M_\odot)\dol/\dls}\,$AU.  Hence, if the event
is observed from a satellite displaced from the Earth in the observer
plane by $d_\sat\sim{\cal O}$(AU),
the event will look significantly different.  It will have a different
$t_0$ and a different $\beta$ (Refsdal 1966; Gould 1994b).  As shown in
Figure \seven(b-c), one can use the difference in the two light curves to 
reconstruct 
the separation in the Einstein ring $\Delta\bx$,
$$\Delta \bx = (\omega\Delta t,\Delta \beta),\eqn\delxdef$$
where $\Delta t$ and $\Delta \beta$ are the differences in the light-curve
parameters as seen from the satellite and Earth.  By measuring $\Delta \bf x$,
one can determine $\tilde r_e$ [see Fig.\ \seven(a)],
$$\tilde r_e = {d_\sat\over \Delta x}.\eqn\retildeeval$$

	In fact, the situation is not quite so simple.  As portrayed in
Figure \seven(b), the source passes on the same side of the lens as seen from
the Earth and the satellite.  In this case $\Delta \beta = \Delta\beta_-$,
where $\Delta\beta_\pm \equiv |\beta'\pm\beta|$ and
$\beta'$ is the impact parameter as seen from the satellite.  But in
fact, from the measurement of $\beta$ and $\beta'$ alone, one has no way
of knowing whether the source is on the same or opposite side as seen by the
two observers.  For the
opposite-side case $\Delta\beta =\Delta\beta_+$.  Hence there are two possible
solutions for 
$\tilde r_e = d_\sat [(\omega\Delta t)^2 + (\Delta\beta_\pm)^2]^{-1/2}$.
Fortunately, this degeneracy can be broken by measuring the fractional 
difference in time scales between the Earth and satellite, 
$\Delta\omega/\omega$, which arises from the relative motion of the Earth
and satellite, $v_\sat$.  The larger the projected speed of the lens, 
$\tilde v\equiv \omega \tilde r_e$, is relative to $v_\sat$, the smaller
$\Delta\omega/\omega$.  Hence, the time-scale difference allows one
to choose the correct solution (Gould 1995a).  Numerical simulations by
Gaudi \& Gould (1996) show that 2\% photometry
(which for bulge giant sources could be obtained using a $\sim 25\,$ cm 
telescope) would be adequate to measure $\tilde r_e$.  The reason for
focusing on giants will become clear in the next section.

	It is possible in principle to use the Earth's orbital motion
to measure parallax from the ground (Gould 1992), and in fact one
such measurement has already been made (Alcock 1995c).  However, for typical
events the Earth does not change its velocity enough over the event time
scale to produce a significant effect.

	The primary motivation for launching a parallax satellite would be
to determine the nature of the MACHOs now
apparently being detected toward the LMC by MACHO and EROS 
(Alcock et al.\ 1993, 1995b, 1995d, 1996a, 1996c; Aubourg et al.\ 1993, 1995;
Ansari et al.\ 1996).  
For these objects, measurement of $\tilde v$ alone
would distinguish among objects in the Galactic disk, Galactic halo, and LMC
(Boutreux \& Gould 1996).  For lensing events seen toward the bulge,
parallax observations do not by themselves provide such a clear cut
distinction among populations, although they do help (Han \& Gould 1995).
However, for the bulge, unlike the LMC, it is possible to obtain a third
piece of information, $\theta_e$, for a representative subset of events.

\section{Measurement of $\theta_e$}

	There are two classes of methods which have been proposed for
measuring $\theta_e$: those that work best when $\theta_e$ is small
and those that work best when $\theta_e$ is large. 

	The first class of methods relies on the finite size of the source.
	If the lens transits or comes sufficiently close to the source, then
the magnification will deviate from the point-source formula \aofx\
(Gould 1994a; Nemiroff \& Wickramasinghe 1994; Witt \& Mao 1994).  This
can be seen most easily by considering the case of $x\rightarrow 0$ for
which $A(x)\rightarrow x^{-1}\rightarrow\infty$.  For simplicity consider
a source with uniform surface brightness, since for these the magnification
is just the area of the image divided by the area of the source.
The image ring in Figure \three(a) appears to have a width about equal to the 
radius of
the star, $\theta_*$.  In fact, one finds from equation \xpdef\ that the
width is exactly $\theta_*$.  Hence the maximum magnification is given by
$$A_\max \sim {2\pi\theta_e\theta_*\over \pi\theta_*^2} = 
2{\theta_e\over\theta_*}.\eqn\amaxeq$$  
By frequently monitoring the event, one can identify the time (and hence the
value of $x=x_*$) when the lens passes over the edge of the star.  The
value of $\theta_*$ can be determined from the star's measured temperature
and Stefan's law.  Hence, one can compute $\theta_e = x_*^{-1}\theta_*$.

	For fixed $\theta_e$ and $\theta_*$, the fraction of events with
such transits is $\theta_*/\theta_e$.  For an ensemble of source stars 
at the Galactic center $(R_0=8\kpc$) with mean physical radius $\VEV{r_*}$,
the fraction is
$${\VEV{\theta_*}\over\theta_e} = {\VEV{r_*}/R_0\over\theta_e} =
{13\mu{\rm as}\over \theta_e}\,{\VEV{r_*}\over 22 r_\odot},\eqn\thratio$$
where the normalization is relative to the mean radius of bulge giants,
$\VEV{r_*}\sim 22r_\odot$ (Gould 1995d) and $r_\odot$ is the solar radius.
Since the distribution of Einstein radii is expected to peak at
$\theta_e\sim 130\,\mu$as (see below), only $\sim 10\%$ of giant-source
events yield measurements of $\theta_e$.  For the more common turn-off stars,
the fraction is $\sim 1\%$.  This is the principal (but not the only) reason 
why lensing searches should be tuned to finding giant-source events
(Gould 1995d).  Finite source effects have been observed in one ``normal''
microlensing event seen toward the bulge (Alcock et al.\ 1996d) as well
as in two binary-lens events, one toward the bulge (Udalski et al.\ 1994c; 
Alcock et al.\ 1996b)
and one toward the LMC (Alcock et al.\ 1996c).

	It is in fact possible to about double the number of measurements
relative to equation \thratio.  If the lens passes close to the limb
of the star but does not actually transit, there is still a noticeable
change in the magnification relative to the point-source formula,
$\delta A/A = \Lambda/8 (x_*/x)^2$.  Here $\Lambda$ is the second radial
moment of the stellar flux normalized so that $\Lambda=1$ for a uniform
disk.  Unfortunately, if the lens does not transit the star, then
this effect simply masquerades as a change in $\beta$, so it cannot
actually be detected (Gould \& Welch 1996).  However, giant stars are
limb darkened by different amounts in different spectral bands.  If the 
star could be resolved, it would have a blue core and a red rim.  
This can give rise to color effects during the microlensing event because
the lens magnifies the near part of the star more than the far part 
(Witt 1995).  These color effects cannot be
mimicked by a change in lensing parameters, since point-source lensing
is achromatic.  For the specific choice of $V$ and $H$ bands, Gould \&
Welch (1996) find $\Lambda^H-\Lambda^V=0.07$, which implies a color shift
of $\sim 0.01(x_*/x)^2$ mag.  By aggressively monitoring giant star events
with an optical/infrared camera equipped with a dichroic beam splitter, they
estimate that $\theta_e$ could be measured for $\beta\leq 2 x_*$.  See
also the related idea of Loeb \& Sasselov (1995) to make use of Ca II
limb brightening.

\topinsert
\mongofigure{ps.thetae}{6.4}{5.5}{6.4}{
Bold solid curve is the expected distribution of Einstein radii $\theta_e$ 
toward the Galactic
bulge based on the best fit power-law model of Han \& Gould (1996a)
(cf.\ Fig.\ \six).  The fraction of giant-source events for which 
$\theta_e$ can be measured by optical/infrared photometry is shown as
a solid curve.  Note that this fraction $\propto\theta_e^{-1}$.  For 
$\theta_e>0.3\,$mas, it is expected that $\theta_e$ can be measured for
$\sim 35\%$ of events with the CHARA interferometer now under construction
(bold dashed curve).  A hypothetical interferometer with 3 times better
resolution but the same magnitude limit is shown by a dashed curve.  
}\endinsert
\FIG\nine{
Bold solid curve is the expected distribution of Einstein radii $\theta_e$ 
toward the Galactic
bulge based on the best fit power-law model of Han \& Gould (1996a)
(cf.\ Fig.\ \six).  The fraction of giant-source events for which 
$\theta_e$ can be measured by optical/infrared photometry is shown as
a solid curve.  Note that this fraction $\propto\theta_e^{-1}$.  For 
$\theta_e>0.3\,$mas, it is expected that $\theta_e$ can be measured for
$\sim 35\%$ of events with the CHARA interferometer now under construction
(bold dashed curve).  A hypothetical interferometer with 3 times better
resolution but the same magnitude limit is shown by a dashed curve.  
}
	Figure \nine\ shows the expected distribution of $\theta_e$ based on
the best-fit model of Han \& Gould (1996a).  The lower solid curve shows
the fraction of giant source events for which measurement of $\theta_e$
is possible with optical/infrared photometry.  The fraction is a respectable
20\% at the peak of the curve, but falls drastically along the long tail
toward larger $\theta_e$.  If measurements are to be made for a representative
sample of events (which is essential to determining the MF), then some
alternative method must be used for large $\theta_e$.  Several such methods
have been proposed: direct imaging (Gould 1992), apparent proper motion
of the mean image position (Hog, Novikov, \& Polnarev 1995), and even lunar
occultations (Han, Narayanan, \& Gould 1996).  The best prospect at the
present time seems to be direct imaging using infrared interferometry.

	The CHARA interferometer (McAlister et al.\ 1994, 1995) now under 
construction is expected
to be able to resolve images separated by $0.6\,(\lambda/1.65\,\mu$m$)\,$mas
at 13th magnitude.  The magnitude limit forces one into the infrared because
few lensing events are this bright in the optical, even when the sources are
restricted to giants.  Bulge giants have $V_0<16$, and in addition generally 
suffer several magnitudes of extinction.  They are intrinsically brighter in
the infrared $(H_0<13.5)$ and in addition the extinction is typically low,
$A_H<0.5$.  For the image splitting to be observed, the {\it fainter} image 
must be brighter than the magnitude 
limit.  Thus, the half of giants with $H<13$ 
require $\beta\lsim 1/2$ and the remainder require $\beta\lsim 1/4$ [cf.\ eq.\
\apm].  I therefore estimate that $\sim 35\%$ of giant-source events with 
image separations $>0.6\,$mas (and hence Einstein rings $\theta_e>0.3\,$mas) 
could yield  measurements of $\theta_e$.
This fraction is shown in Figure \nine.

\section{Prospects}

	The instruments that would be required to make microlensing  
measurements of the MF are already being built or are the subject of active 
proposals.  A parallax satellite was proposed as a MIDEX project, although
it failed to win approval in the current round.  
	The CHARA interferometer is scheduled for completion in 1999
(H.\ McAlister 1996, private communication).
D.\ L.\ DePoy (1996, private communication) has designed an optical/infrared
camera and a proposal to build it is now pending.  Ultimately, one
would want to place one of these cameras on each of several 1m class telescopes
around the world in order to obtain 24 hour coverage of ongoing events.
Two groups (PLANET and GMAN) are already monitoring ongoing 
microlensing events with conventional optical telescopes from sites 
encircling the globe.

	From Figure \nine\ it is clear that even with all this equipment in 
place it would be possible to measure masses for only $\sim 15\%$ of 
giant-star events.  However, interferometric technology is advancing at
rapid rate.  A factor $\sim 3$ improvement in resolution would allow one
to measure the lens masses for $\sim 35\%$ of events and moreover would
produce a representative sample of lenses, that is one with only modest 
(and hence easily corrected) selection bias (see Fig.\ \nine).  

\chapter{Mass Function of the LMC}

	As discussed in the introduction, the microlensing observations 
toward the LMC are aimed primarily at discovering MACHOs in the Galactic
halo.  Inevitably, however, there will be some microlensing by stars in
the Galactic disk, Galactic spheroid (stellar halo), and the LMC itself.
To date, about eight candidate events have been reported by MACHO and two by
EROS.  Assuming that all these events are real, this rate implies
an ``optical depth'' (probability that a given star is lensed at any
given time) of $\tau\sim 2\times 10^{-7}$.  Star counts place strong
constraints on the contribution to $\tau$ from luminous stars in the 
Galactic disk (Gould et al.\ 1996) and spheroid (Mould 1996; Flynn, Gould, 
\& Bahcall 1996).  On the other hand, Sahu (1994a,b) has argued that the
majority of the events come from LMC stars.  Although this is contested
(Gould 1995b), one of the eight MACHO candidates is almost certainly
in the LMC (Alcock et al.\ 1996c).  The reason this is known is that the 
event is a binary lens with two ``caustic crossings''.  For such events
it is possible to measure $\theta_e$ using a variant of the method discussed
in \S\ 5.2.  One can then determine the proper motion $\mu=\omega\theta_e$
which turns out to be consistent with the low values expected for LMC lenses,
but is much too low for closer lenses.  

	To use  microlensing to measure the LMC MF, one must
first isolate a subsample of all events which are due to LMC lenses.  One
could do this with satellite observations of the type described in \S\ 5.1.
For LMC stars with e.g., $M\sim 0.3\,M_\odot$, $\dls\sim 1\,$kpc, one
finds $\tilde r_e\sim 75\,$AU.  Hence, LMC events would be recognized as
those with $\Delta x = d_\sat/\tilde r_e\ll 1$.  However, given reasonable
satellite design parameters (e.g., Boutreux \& Gould 1996), it is likely
that $\Delta x$ would only be shown to be consistent with zero, not actually
measured.  Thus there would be no additional information from the parallax
measurement beyond that the lens was in the LMC. Thus, the measurement of
the MF is substantially more difficult for the LMC than the bulge,
not only because there is a lower event rate, but because there is less
information for each event measured.

\chapter{Mass Function of M31 from Pixel Lensing}
	
	It has proven difficult to measure the MF even in the solar 
neighborhood, so the possibility of measuring the MF of the M31 bulge
might seem remote at best.  In fact, rapid progress is already being made. 

	The problem with doing lensing experiments toward the M31 bulge
is that none of the stars are resolved.  Instead, each pixel contains
dozens or hundreds of bright stars and it is therefore impossible to measure 
the individual
flux from any one of them.  Recall from equation \foft, however, that what
is of interest is not the flux from the star, but the difference in flux
from its unmagnified state.  Crotts (1992) and Baillon et al.\ (1993) pointed
out that this can be measured by monitoring the time dependence of the 
pixel counts.  Both groups have initiated pixel lensing searches toward
M31 and have achieved impressive initial results.

	To understand pixel lensing, consider two images taken in identical
seeing, with identical exposure times and atmospheric extinction, and
perfectly aligned.  One image is taken before and one during the lensing event.
If the two images are subtracted, the difference image will look like a blank
patch of sky containing a single stellar point spread function (PSF),
with total flux $F=F_0(A-1)$, exactly as in equation \foft.  One then 
constructs a light curve just as one would do in ordinary ``classical'' 
lensing.

\topinsert
\mongofigure{ps.newref}{6.4}{5.5}{6.4}{
Detection of variable stars in M31 using pixel-lensing techniques
(Tomaney \& Crotts 1996). 
(a) An $R$ band reference image of M31 
is constructed from 20 800 s integrations taken in $\sim 1''$ seeing
between 20 Nov 1995 and 3 Dec 1995.  The field is an $85''$ (256 pixel) 
square which is $400''$ from the center of M31 along the far minor axis. It
has mean surface brightness $R\simeq 20\,$mag arcsec${}^{-2}$.  The
large scale gradient has been removed in this display.  Most of the
variation is due to surface brightness fluctuations rather than individual
stars. The bright object at the far right and $\sim 20''$ from the bottom
is probably a globular
cluster.  
From work to be published by Tomaney \& Crotts (1996).  
Reproduced with permission.
}\endinsert
\FIG\ten{
Detection of variable stars in M31 using pixel-lensing techniques
(Tomaney \& Crotts 1996).  On top, an $R$ band reference image of M31 
is constructed from 20 800 s integrations taken in $\sim 1''$ seeing
between 20 Nov 1995 and 3 Dec 1995.  The field is an $85''$ (256 pixel) 
square which is $400''$ from the center of M31 along the far minor axis. It
has mean surface brightness $R\simeq 20\,$mag arcsec${}^{-2}$.  The
large scale gradient has been removed in this display.  Most of the
variation is due to surface brightness fluctuations rather than individual
stars. The bright object at the far right and $\sim 20''$ from the bottom
is probably a globular
cluster.  On bottom, the reference image (a) is subtracted from an
image constructed from 10 800 s exposures taken on 21 Oct 1995 in 
$1.\hskip-2pt ''2$
seeing after photometric and geometric registration and after convolving the
reference image to bring it to the same seeing.  Images (a) and (b) are
shown at the same stretch.  The bright and dark circles are PSF-shaped
residual fluxes which are positive and negative respectively.  The
brightest flux differences correspond to $R\sim 21$ and the faintest to
$R\sim 23$.  While most of the objects seen in this difference image are
undoubtedly variable stars, the image demonstrates that the pixel-lensing
technique is viable and should be able to detect variation due to microlensing.
From work to be published by Tomaney \& Crotts (1996).  
Reproduced with permission.
}
\global\advance\mongocount by -1\bigskip
\topinsert
\mongofigure{ps.dif}{6.4}{5.5}{6.4}{
(b): 
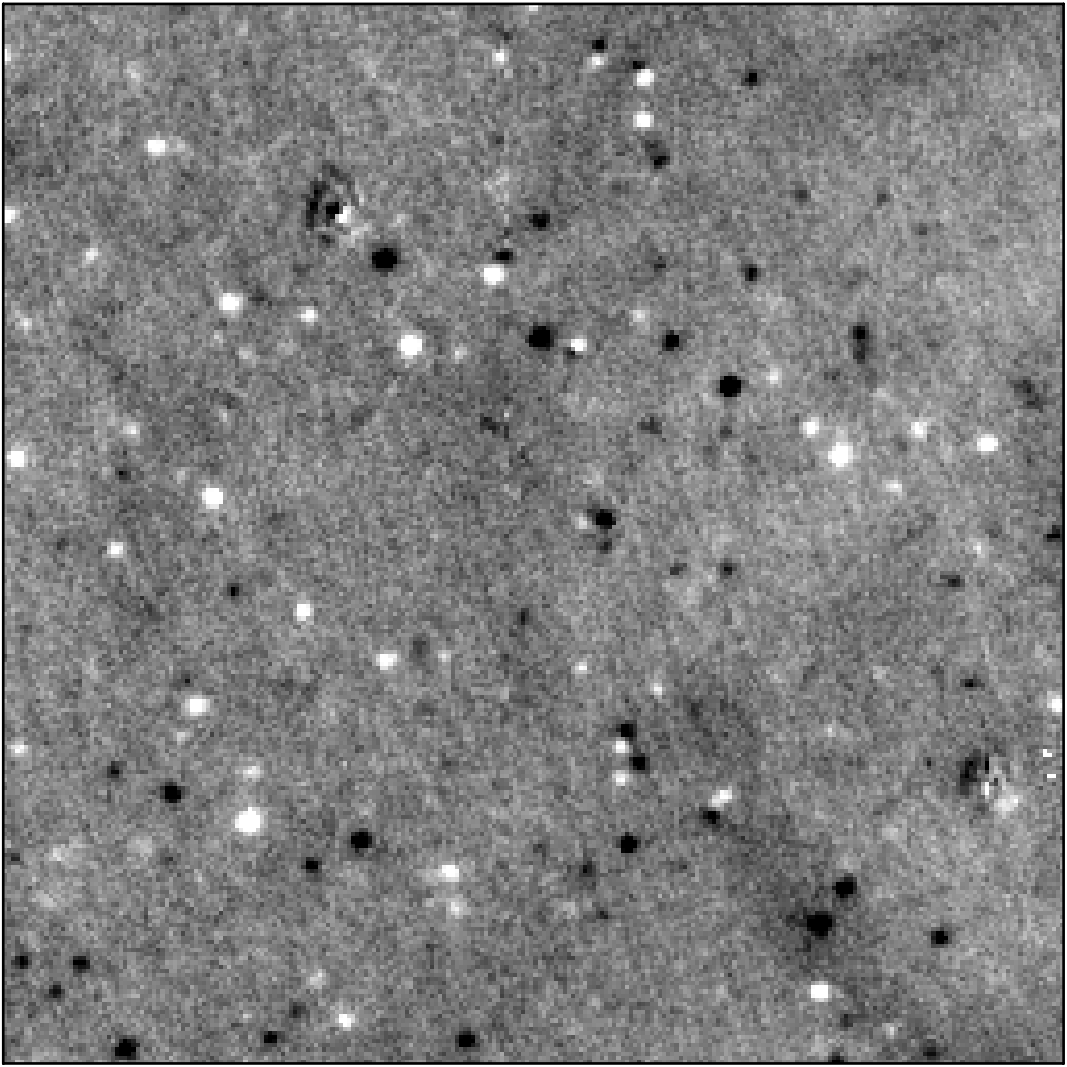
The residuals formed by subtracting the reference image 7(a) from an
image constructed from 10 800 s exposures taken on 21 Oct 1995 in 
$1.\hskip-2pt ''2$
seeing after photometric and geometric registration and after convolving the
reference image to bring it to the same seeing.  Images (a) and (b) are
shown at the same stretch.  The bright and dark circles are PSF-shaped
residual fluxes which are positive and negative respectively.  The
brightest flux differences correspond to $R\sim 21$ and the faintest to
$R\sim 23$.  While most of the objects seen in this difference image are
undoubtedly variable stars, the image demonstrates that the pixel-lensing
technique is viable and should be able to detect variation due to microlensing.
From work to be published by Tomaney \& Crotts (1996). 
Reproduced with permission.
}\endinsert
	Of course the seeing et cetera is never identical, so pixel lensing
seems to many to be a hopeless illusion.  However, it is possible to convolve
images to the same seeing, to align them photometrically and geometrically,
and to obtain difference images very much like those described.  Both
groups have used variations on this technique to find large numbers of
variable stars.  In particular, Tomaney \& Crotts (1996) have produced a
spectacular difference image (reproduced as Fig.\ \ten)
in which the variables appear as positive and
negative PSFs.  It is not yet possible to confirm microlensing in these
studies because of an inadequate baseline, but it should be possible in the
near future.

	What can be learned about the M31 MF from pixel lensing?  For the
foreseeable future, the only information available for most events will
be the time scale, $t_e$.  Even the time scale can be difficult to recover
for those events in which a very faint star is highly magnified $(\beta\ll 1)$.
In this case,
$$F= F_0(A-1)\rightarrow {F_\max\over[1 +(t-t_0)^2/t_\eff^2]^{1/2}},\eqn\fmax$$
where $F_\max \equiv F_0/\beta$ and $t_\eff\equiv \beta t_e$.  Since $t_e$
does not appear in the limiting form \fmax\ of the flux, it cannot generally
be measured for these events.  Nevertheless, if the signal-to noise ratio of
the event is sufficiently high $(S/N\gsim 80)$, then there is enough 
information to distinguish equation \foft\ from its limiting form \fmax.
In these cases, it is possible to measure the time scale, just as it is for
Galactic events (Gould 1996).  Such high $S/N$ events can occur in M31 either 
because the source star is relatively bright or the impact parameter $\beta$
is small (Colley 1995; Han 1996).  Han (1996) estimates that
$t_e$ can be measured for
$\sim 10$ events per year assuming a modest dedicated experiment.
Thus, pixel lensing will allow a statistical
determination of the M31 MF of the type described in \S\ 4.  In particular,
one can test whether the M31 bulge MF is consistent with the more accurately
determined MF of the Galactic bulge.

	What are the prospects for measuring $\theta_e$ and $\tilde r_e$ for
M31 lenses?
Han \& Gould (1996b) find that $\theta_e$ can be measured from finite-size
effects (cf.\ \S\ 5.2) for $\lsim 1$ event per year.  Moreover, 
$\tilde r_e\sim 10^3\,$AU, so parallax measurements are out of the question 
for the near term.  

	In addition to obtaining information on the M31 MF, one would
also find out about the LF.  Recall from equation \foft\ that a microlensing
light curve is a function of 4 parameters, including the source flux $F_0$.
This quantity is then measured in all events with good enough signal to noise
to determine the time scale, estimated above at $\sim 10$ per year.
Normally one does not think of microlensing
as a way to determine the brightness of stars because it is so much easier 
to directly measure their flux.  However, the stars in the M31 bulge are
not resolved, so pixel lensing may be the only way to measure their flux
and so determine the LF (Gould 1996).

\chapter{Pixel Lensing of M87}

	There are several good reasons to search for microlensing in M87, the
giant elliptical in the core of the Virgo cluster.
First, a significant fraction of the halo of the Virgo cluster might well
be made of MACHOs.
Preliminary results of microlensing searches toward the LMC seem to
indicate that $\gsim 20\%$ of the dark halo inside the LMC is composed
of MACHOs 
(Alcock et al.\ 1993, 1995b, 1995d, 1996a, 1996c; Aubourg et al.\ 1993, 1995,
Ansari et al.\ 1996).  That is, the baryonic component
of the Milky Way halo appears to be at least as massive as the visible
baryons in the disk and bulge.  One might imagine that half the available
baryons formed MACHOs early in the Galaxy's history and the other half
settled into a proto-disk and proto-bulge.  Now consider a Milky Way-like
galaxy forming on the edge of the Virgo cluster.  It has time to form
its MACHOs, but before the proto-disk can start forming stars, the galaxy
falls into the cluster center and is stripped of its remaining gas by
the intra-cluster medium.  The MACHOs would then continue in their orbits,
either as a relatively coherent dark galaxy or as a diffuse collection of
Virgo halo objects.  Either way, these objects would give rise to
pixel lensing events of M87 (Gould 1995f).
Second, M87 is a giant elliptical galaxy, and its stellar population could
differ radically from those of the Galaxy and M31.  Pixel lensing could
test for such differences.

	Unfortunately, pixel lensing increases in difficulty as the fourth 
power of the
distance of the target galaxy.  Two identical galaxies at different
distances, $d_1$ and $d_2$, will have the same surface brightness $S$.  
If observed from the same telescope, a resolution element $\Omega_\psf$ will 
contain the same amount of flux, $S\Omega_\psf$.  Thus the photon noise
will be the same for measurements of the same duration, $\Delta t$.  On
the other hand, the flux from a typical lensing event will be lower by
a factor $(d_2/d_1)^{-2}$.  Hence the ratio of the signal to noise (S/N)
for the two galaxies is 
$${\rm (S/N)_2\over (S/N)_1}=\biggl({\Delta t_2\over \Delta t_1}\biggr)^{1/2}
\biggl({d_2\over d_1}\biggr)^{-2}.\eqn\snratios$$
This implies that to reach more distant galaxies requires longer exposure times
by $\Delta t\propto d^4$, or telescopes with proportionately larger apertures.

	Another approach is increased resolution.  Dedicated observations
with {\it HST} would yield $\sim 18 f$ Virgo cluster halo events and
another $\sim 3$ events from M87 stars per day (Gould 1995f).  Here $f$ is the 
fraction of the Virgo halo composed on MACHOs.  The two populations could
be distinguished, at least statistically, because they would have different
angular distributions.

\chapter{Star Formation History of the Universe}

	Only a few years ago, measurements of the MF were restricted to stars 
in the ``solar neighborhood'', the nearest 10s of pc.  Today, it is already
possible to probe not only distant parts of our own Galaxy, but even M87 
at 10s of Mpc.  Nevertheless, from a cosmological point of view the Virgo
cluster is still the ``solar neighborhood''.  In Table 1, M87's redshift
$z$ is still written in scientific notation.  Will it ever be possible
to determine the MF at such large distances that they represent different
epochs in the history of the universe?
\midinsert
\overfullrule=0pt
$$\vbox{\halign{#\hfil\quad& #\hfil\quad&\hfil#\hfil\cr
\multispan{3}{\hfil TABLE 1 \hfil}\cr
\noalign{\bigskip}
\multispan{3}{\hfil Summary of Probes of The Stellar Mass Function\hfil}\cr
\noalign{\medskip}
\noalign{\hrule}
\noalign{\smallskip}
\noalign{\hrule}
\noalign{\smallskip}
\hfil Method\hfil&\hfil Star Field \hfil&\hfil$z$\hfil\cr
\noalign{\smallskip}
\noalign{\hrule}
\noalign{\smallskip}
Parallax & 10 pc from Sun & $2\times 10^{-9}$\cr
\noalign{\smallskip}
Ground Photometry & 100 pc from Sun & $2\times 10^{-8}$\cr
\noalign{\smallskip}
HST Photometry & 3 kpc from Sun & $7\times 10^{-7}$\cr
\noalign{\smallskip}
Microlensing & MW Bulge and Disk & $2\times 10^{-6}$\cr
\noalign{\smallskip}
Pixel Lensing & M31 Bulge & $2\times 10^{-4}$\cr
\noalign{\smallskip}
Pixel Lensing & M87 & $4\times 10^{-3}$\cr
\noalign{\smallskip}
\noalign{\hrule}
}}
$$

\endinsert

	Actually, this may not be so difficult as it would first appear.
There are 100 quasars to $B=22$ per square degree.  One could monitor
these for gravitational lensing over say 10,000$\,\rm deg^2$.  That
would be $10^6$ quasars.  The rate, $\Gamma$, at which known stars in 
galaxies would generate events is
$\Gamma\sim 20\,\rm yr^{-1}$.  If the density of MACHOs (in or out
of galaxies) were $\Omega_{\rm MACHOs}\sim 1\%$, these would generate
an additional $\Gamma\sim 200\,\rm yr^{-1}$.  The stellar events would
typically last $\sim 10\,\rm yr$, while the MACHO events would last
$\sim 3\,\rm yr$  (Gould 1995e).  
Because the events do last several years,
it would be sufficient to monitor the quasars only several times per year.
In fact it would be unnecessary even to identify the $10^6$ quasars.  The
10,000 square degrees could be monitored with say a 4 square degree field
on a 1 or 2 m telescope.  Lensed quasars would be found by the same technique
used in pixel lensing searches (see Fig.\ \ten).  The visible disks of galaxies
cover about 1\% of the sky and hence 1\% all quasars lie behind these disks 
(Gould 1995c).  If quasars in this subsample could be 
identified, it might be advantageous to monitor them more carefully.

	What could be learned about stars in such a survey?  First, the
star formation history of the universe.  If most stars formed before the
epoch of quasars, then the optical depth to $z=4$ quasars would be much
higher than to $z=1$ quasars, because the path length to the former would
be densely populated with stars.  On the other hand, if most stars formed at
$z<1$, then the distant quasars would have only a slight advantage in
optical depth (Gould 1995e).  

	Second, one could learn, at least statistically, 
about the MF.  Measuring the mass of individual lensing
stars would be more difficult, although even this would be possible given
a sufficient investment.  For most stars, or at least the stars in
galaxies, one could identify the host galaxy by following up the detection
with deep imaging.  Once the redshift of the galaxy were measured, one
would have two measurements (distance and time-scale of the event) against
three unknowns (the mass, distance, and transverse speed).  In other words,
the mass could be statistically estimated by assuming a distribution of 
transverse speeds of galaxies.  It is in fact possible to measure the
scale of the transverse speeds from the monitoring experiment by comparing
the lensing rate toward quasars in directions parallel and perpendicular
to the Sun's motion relative to the microwave background.  But it is
also possible to measure the transverse velocities of the lensing stars
themselves, provided that one observes the event simultaneously
from a satellite in a Neptune-like orbit (Gould 1995c).  Such parallax
measurements would be very similar to the measurements 
proposed
for Galactic lensing events (Gould 1994b, 1995a) but on a much larger scale.

	In fact, the first confirmed microlensing by a star was discovered 
not in our Galaxy but in Huchra's Lens (in front of qso 2237+0305) 
(Irwin et al.\ 1989;
Corrigan et al.\ 1991).  While only at $z=0.04$, this serendipitous 
discovery may one day be regarded as the first step toward a measurement
of the star-formation history of the universe.

	One might wonder how
one could distinguish genuine quasar microlensing events from ordinary
quasar variations.  In fact, this is not difficult in principle: spectroscopic
follow up would reveal a light echo in the broad line region of 
any variation coming from the continuum.  No such effect would be seen for
microlensing.  Unfortunately, the telescope time necessary to follow up
all variations among $10^6$ quasars would be prohibitive.  However, in the
original survey, one could identify a subset of relatively quiescent quasars
or, more to the point, eliminate quasars that showed too much variation.
Whether such a population of quiescent quasars exists is not presently known,
but is likely to be determined as a by-product of large scale supernova 
searches to be carried out over the next few years.

	I should add that the various proposals envisaged in \S\S\ 8 and 9 
differ dramatically in terms of resource requirements.  A dedicated
quasar variability search is the same order of project as the MACHO searches
that are now being undertaken by several groups.
Dedicated {\it HST} observations of M87 would be similar in scope to the
Hubble Deep Field (HDF) observations and hence would require either 
Director's Discretionary Time (as with HDF) or at least very broad community
support.  On the other hand, a 1 m telescope in a Neptune-like orbit would
be on the general scale of the most ambitious space-science projects, like
{\it HST} itself.

\chapter{Planet Searches}

	Microlensing can be a powerful tool for studying planets as well
as stars.  Once a microlensing event is under way, one can monitor the event
to search for the short $(\lsim 1\,$day) perturbations characteristic of
a planet.  This in fact is the primary motivation of the two lensing follow up
networks, PLANET (Albrow et al.\ 1996) and GMAN (Pratt et al.\ 1996).

	Microlensing searches for planets were first suggested by Mao \&
Paczy\'nski (1991) who analyzed these events as the low mass limit of
binary-lens microlensing.  Binary microlensing is far more
complicated than point-mass microlensing.  In particular, the light curves
do not in general resemble a superposition of two point-mass curves of the
type described by equation \aofx.

	Planetary lensing is somewhat simpler than binary lensing because
it can be described by a fourth order equation which can be solved 
analytically, rather than a fifth order equation which cannot (Gould \&
Loeb 1992).  Nevertheless, a full treatment of planetary lensing is beyond
the scope of the present review.  For a large fraction of planetary
lensing events, however, the light curve does in fact qualitatively resemble
a superposition of two light curves.  The primary light curve has time scale
$t_e$ and the perturbing light curve has duration $t_p\sim (m/M)^{1/2}t_e$,
where $M$ and $m$ are the masses of the lensing star and its planet 
respectively.  The planetary perturbation is shorter than the main event
because the planet has a proportionately smaller Einstein ring, $\theta_p
=(m/M)^{1/2}\theta_e$.  This picture, while considerably oversimplified,
does permit one to understand what can be learned from planetary microlensing
events. 

	One must distinguish between two classes of planetary lensing events,
depending on whether the planet Einstein radius $\theta_p$ is bigger or
smaller than the stellar radius $\theta_*$.  For an average giant source
($\theta_* = 22 r_\odot/R_0)$, and a planetary system at $\dol\sim 6\,$kpc
toward the bulge, the boundary between these regimes is at $m\sim 85\,M_\oplus$
where $M_\oplus$ is the mass of the Earth.  

	Consider first the case of $\theta_p>\theta_*$, the ``Jupiters''.
Naively, the fraction of events where the planet causes a recognizable
perturbation is $\sim \theta_p/\theta_e=(m/M)^{1/2}$, or $\sim 3\%$ for
a Jupiter in orbit around a solar type star.  In fact, a more detailed analysis
shows that the lensing star actually stretches the Einstein ring of the
planet so that the fraction of noticeable events is much higher, $\sim 17\%$
for a Jupiter (Gould \& Loeb 1992).  One can measure two quantities in this
case, the ratio of the masses of the planet to the star and the projected
separation of the planet from the star in units of the Einstein ring.  
The mass ratio
is determined from the measured ratio of time scales $m/M = (t_p/t_e)^{2}$,
and the projected separation in units of the Einstein ring is simply the
value of $x$ when the planetary event occurs.  Since the mass and
Einstein radius of the parent star can be estimated to within a factor
of $\sim 2$--3, the mass and separation of the planet are known equally
well.  If there is additional information such as $\tilde r_e$ from
parallax (\S\ 5.1), $\theta_e$ (\S\ 5.2), or a measurement of the spectral
type of the parent star (Gould \& Loeb 1992;  Buchalter, Kamionkowski, \& Rich
1996), then the mass and separation can be further constrained or precisely
specified.

	The case of $\theta_p<\theta_*$ is rather different.  First, the
probability of a planetary event scales not with $\theta_p$ but
with $\theta_*$.  This means that no matter how small the planet, there
is a probability $\sim\theta_*/\theta_e$ that a planetary event will take
place (assuming of course that the star has a planet in the neighborhood
of the Einstein ring).  For an $M=0.5\,M_\odot$ star at 6 kpc lensing an
average giant, this is $\sim 3\%$, small but not negligible given that
an aggressive search toward the bulge could detect $\sim 100$ giant-source 
lensing events per year (Gould 1995d).  
In fact, microlensing is the only known technique
for detecting Earth-mass planets from the ground.  Note that the sensitivity
of the survey could be increased somewhat by also monitoring the more numerous
(but much smaller) turn off stars in the bulge.

	However, a second major effect when $\theta_p<\theta_*$ is that
the maximum size of the perturbation is only $\sim 2(\theta_p/\theta_*)^2$,
or $\sim 1.4\%$ for an Earth-mass planet with the above geometry.  The
events would last for the time it took for the planet to cross the source
star, typically $\sim 10\,$ hours.  It is not difficult to detect $\sim 1\%$
variation of a bright giant which is monitored continuously.  However, since
the detections are not repeatable, one might well question whether any
particular event was real, or whether it resulted from intrinsic variation
or just an instrument glitch.  In order to confirm events on a routine basis,
it is therefore important to make simultaneous optical/infrared images
using a dichroic beam splitter such as has been proposed by D.\ L.\ 
DePoy (1996 
private communication).  Not only would such observations have the important
feature of redundancy, but limb darkening would generate a characteristic 
color variation during the planetary perturbation which would unambiguously
distinguish it from any other possible cause (Gould \& Welch 1996).

	A third feature of small-mass planet events is that it is almost
always possible to measure $\theta_p$, because the planet almost always
transits the image of the star.  This means that the mass is generally
better constrained for these events.  Moreover, if there is a parallax
measurement, the mass can be determined precisely.

\chapter{Summary}

	The original object of microlensing surveys was to find the dark
matter in the halo of the Milky Way (Paczy\'nski 1986) and other galaxies
(Crotts 1992; Baillon et al.\ 1993; Gould 1993).  However, even before
the surveys got started, they began to expand their horizons to become
probes of visible as well as dark matter
(Paczy\'nski 1991; Griest et al.\ 1991).  Once underway, the successes
of the microlensing surveys inspired many ideas for extracting additional
information from the data, in particular methods for constraining or
measuring the masses of the lenses.  With the development of these new
techniques, microlensing promises to provide a powerful probe of the
stellar MF and a method for mapping out the distribution of sub-stellar
objects including planets.  Unlike traditional approaches to measuring
the MF, microlensing is not restricted to the solar neighborhood, or
even the Galaxy, Local Group, or Local Supercluster.  Microlensing
is already giving hints of an unexpected population of low mass stars in
our Galaxy, and may yet tell us something about stars at the edge of the
universe.

{\bf Acknowledgements}:  I would like to thank K.\ Griest, G.\ Newsom, 
J.\ Rich and K.\ Sahu for making many valuable suggestions and to extend 
special thanks to A.\ Tomaney
and A.\ Crotts for giving permission to use their spectacular difference 
image of M31 in advance of publication. 
This work was supported in part by grant AST 94-20746 from the NSF.
\endpage
\Ref\alard{Alard, C.\ 1996, in Proc. IAU Symp.\ 173 (Eds.\ C.\ S.\ Kochanek, 
J.\ N.\ Hewitt), in press (Kluwer Academic Publishers)}
\Ref\amg{Alard, C., Mao, S., \& Guibert, J.\ 1995, A\&A, L17}
\Ref\albrow{Albrow, M., et al.\ 1996, in Proc. IAU Symp.\ 173 
(Eds.\ C.\ S.\ Kochanek, 
J.\ N.\ Hewitt), in press (Kluwer Academic Publishers)}
\Ref\Alcock{Alcock, C., et al.\ 1993, Nature, 365, 621}
\Ref\Alcock{Alcock, C., et al.\ 1995a, ApJ, 445, 133}
\Ref\Alcock{Alcock, C., et al.\ 1995b, ApJ, 449, 28}
\Ref\Alcock{Alcock, C., et al.\ 1995c, ApJ, 454, L125}
\Ref\Alcock{Alcock, C., et al.\ 1995d, Phys.\ Rev.\ Lett., 74, 286}
\Ref\Alcock{Alcock, C., et al.\ 1996a, ApJ, submitted}
\Ref\Alcock{Alcock, C., et al.\ 1996b, ApJ, submitted}
\Ref\Alcock{Alcock, C., et al.\ 1996c, ApJ, in preparation}
\Ref\Alcock{Alcock, C., et al.\ 1996d, ApJ, in preparation}
\Ref\ansari{Ansari, R., et al.\ 1996, A\&A, in press}
\Ref\Aubourg{Aubourg, E., et al.\ 1993, Nature, 365, 623}
\Ref\Aubourg{Aubourg, E., et al.\ 1995, A\&A, 301, 1}
\Ref\Bdmo{Bahcall, J.\ N., Flynn, C., Gould, A.\ \& Kirhakos, S.\ 1994, ApJ,
435, L51}
\Ref\bail{Baillon, P., Bouquet, A., Giraud-H\'eraud, Y., \& Kaplan, J.\ 1993,
A\&A, 277, 1}
\Ref\bes{Bessel, M.\ S.\ \& Stringfellow, G.\ S.\ 1993, ARAA, 433}
\Ref\bg{Boutreux, T., \& Gould, A.\ 1996, ApJ, 462, 000}
\Ref\bk{Buchalter, A., Kamionkowski, M., \& Rich, R.\ M.\ 1996, ApJ, submitted}
\Ref\col{Colley, W.\ N.\ 1995, AJ, 109, 440}
\Ref\cor{Corrigan, R.\ T.,\ et al.\ 1991, AJ, 102, 34}
\Ref\crotts{Crotts, A.\ P.\ S.\ 1992, ApJ, 399, L43}
\Ref\flynn{Flynn, C., Gould, A., \& Bahcall, J.\ N.\ 1996, ApJL, submitted}
\Ref\gaudi{Gaudi, B.\ S., \& Gould, A.\ 1996, ApJ, submitted}
\Ref\gone{Gould, A.\ 1992, ApJ, 392, 442}
\Ref\gtwo{Gould, A.\ 1993, ApJ, 404, 451}
\Ref\gtwo{Gould, A.\ 1994a, ApJ, 421, L71}
\Ref\gtwo{Gould, A.\ 1994b, ApJ, 421, L75}
\Ref\gtwo{Gould, A.\ 1995a, ApJ, 441, L21}
\Ref\gtwo{Gould, A.\ 1995b, ApJ, 441, 77}
\Ref\gtwo{Gould, A.\ 1995c, ApJ, 444, 556}
\Ref\gtwo{Gould, A.\ 1995d, ApJ, 447, 491}
\Ref\gtwo{Gould, A.\ 1995e, ApJ, 455, 37}
\Ref\gtwo{Gould, A.\ 1995f, ApJ, 455, 44}
\Ref\gtwo{Gould, A.\ 1996, ApJ, submitted}
\Ref\gthree{Gould, A., Bahcall, J.\ N.\ \& Flynn, C.\ 1996, ApJ, 465, 000}
\Ref\gtwo{Gould, A., \& Loeb, A.\ 1992, ApJ, 396, 104}
\Ref\gtwo{Gould, A., \& Welch, R.\ L.\ 1996, ApJ, 464, 000}
\Ref\grie{Griest, K.\ et al.\ 1991, ApJ, 372, L79}
\Ref\hg{Han, C.\ 1996, ApJ, submitted} 
\Ref\hg{Han, C., \& Gould, A.\ 1995, ApJ, 447, 53} 
\Ref\hg{Han, C., \& Gould, A.\ 1996a, ApJ, 467, 000} 
\Ref\hg{Han, C., \& Gould, A.\ 1996b, ApJ, submitted} 
\Ref\hg{Han, C., Narayanan, V.\ K., \& Gould, A.\ 1996, ApJ, 461, 587} 
\Ref\hmc{Henry, T.\ J.\ \& McCarthy, D.\ W.\ 1993, AJ, 106, 773}
\Ref\hnp{Hog, E., Novikov, I.\ D., \& Polnarev, A.\ G.\ 1995, A\&A, 294, 287}
\Ref\irw{Irwin, M.\ J., Webster, R.\ L., Hewett, P.\ C.,
Corrigan, R.\ T., \& Jedrzejewski, R.\ I.\ 1989, AJ, 98, 1989}
\Ref\jet{Jetzer, P.\ 1994, ApJ, 432, L43}
\Ref\kir{Kiraga, M.\ \& Paczy\'nski, B.\ 1994, ApJ, 430, 101}
\Ref\kirk{Kirkpatrick J.\ D., McGraw, J.\ T., Hess, T.\ R., Liebert, J.\
\& McCarthy, D.\ W., Jr.\ 1994, ApJS, 94, 749}
\Ref\Lio{Liouville, J.\ 1837, Journal de Math\'ematiques Pures et 
Appliqu\'ees, 2, 16}
\Ref\ls{Loeb, A.\ \& Sasselov, D.\ 1995, ApJ, 449, L33}
\Ref\mp{Mao, S.\ \& Paczy\'nski, B.\ 1991, ApJ, 388, L45}
\Ref\mcal{McAlister, H.\ A.\ et al.\ 1994, SPIE, 2200, 129}
\Ref\mcali{McAlister, H.\ A., Bagnuolo, W.\ G., ten Brummelaar, T.,
Hartkopf, W.\ I., \& Mason, B.\ D.\ 1995, The CHARA Array as an ASEPS 
Resource, Technical Report No. 18, 
(Center for High Angular Resolution Astronomy, Georgia State University: 
Atlanta)}
\Ref\mtw{Misner, C.\ W., Thorne, K.\ S.,\& Wheeler, J.\ A.\ 1973, Gravitation,
(San Francisco: Freeman)}
\Ref\mould{Mould, J.\ 1996, PASP, 108, 35}
\Ref\nem{Nemiroff, R.\ J.\ \& Wickramasinghe, W.\ A.\ D.\ T.\ 1994, ApJ, 424, 
L21}
\Ref\Pac{Paczy\'nski, B.\ 1986, ApJ, 304, 1}
\Ref\Pac{Paczy\'nski, B.\ 1991, ApJ, 371, L63}
\Ref\Pac{Paczy\'nski, B.\ 1996, ARAA, submitted}
\Ref\Pratt{Pratt, M., et al.\ 1996, in Proc. IAU Symp.\ 173 
(Eds.\ C.\ S.\ Kochanek, 
J.\ N.\ Hewitt), in press (Kluwer Academic Publishers)}
\Ref\refsdal{Refsdal, S.\ 1966, MNRAS, 134, 315}
\Ref\rei{Reid, I.\ N.,  Hawley, S.\ L., \& Gizis, J.\ E.\ 1995, AJ, 110, 1838}
\Ref\mou{Reid, I.\ N., Tinney, C.\ G., \& Mould J.\ 1994, AJ, 108, 1456}
\Ref\rm{Roulet, R.\ \& Mollerach, S.\ 1996, Physics Reports, submitted}
\Ref\sahua{Sahu, K.\ C.\ 1994a, Nature, 370, 275}
\Ref\sahub{Sahu, K.\ C.\ 1994b, PASP, 106, 942}
\Ref\sca{Scalo, J.\ M.\ 1986, Fund.\ Cos.\ Phys., 11, 1}
\Ref\sef{Schneider, P., Ehlers, J., \& Falco, E.\ E.\ 1992, Gravitational 
Lenses (Berlin: Springer-Verlag)}
\Ref\stobie{Stobie, R.\ S., Ishida, K., \& Peacock, J.\ A.\ 1989, MNRAS, 
238, 709}
\Ref\tomcro{Tomaney, A., \& Crotts, A.\ P.\ S.\ 1996, in preparation}
\Ref\oglea{Udalski, A., et al.\ 
1994a, Acta Astronomica 44, 165}
\Ref\ogleb{Udalski, A., Szyma\'nski, M.,  
Kaluzny, J., Kubiak, M., W., Mateo, M., \& Krzemi\'nski, W.\  
1994b, ApJ 426, L69}
\Ref\oglec{Udalski, A., et al.\ 
1994c, ApaJ 436, L103} 
\Ref\WJK{Wielen, R., Jahreiss, H., \& Kr\"uger, R.\ 1983, IAU Coll.\ 76:
Nearby Stars and
the Stellar Luminosity Function, A.\ G.\ D.\ Philip and A.\ R.\ Upgren eds.,
p 163}
\Ref\witt{Witt, H.\ 1995, ApJ, 449, 42}
\Ref\wm{Witt, H., \& Mao, S.\ 1994, ApJ, 430, 505}
\Ref\zsr{Zhao, H.\ S., Spergel, D.\ N., \& Rich, R.\ M.\ 1995, ApJ, 440, L13}
\refout
\endpage
\end